\documentclass[twocolumn,showpacs,preprintnumbers,amsmath,amssymb,floatfix,superscriptaddress]{revtex4}

\usepackage{epsfig}
\usepackage{graphicx} 
\usepackage{dcolumn}  
\usepackage{bm}       
\usepackage{color}
\usepackage{epstopdf}

\newcommand{\be}{\begin{equation}}
\newcommand{\ee}{\end{equation}}

\newcommand{\paper}{paper}
\newcommand{\tr}{\textrm{tr}}

\begin{document}

\title{Detection and engineering of spatial mode entanglement with
  ultra-cold bosons}

\author{J. Goold} \email{jgoold@phys.ucc.ie}

\affiliation{Department of Physics, University College Cork, Cork,
  Republic of Ireland} 

\author{Libby Heaney} \email{ l.heaney1@physics.ox.ac.uk}

\affiliation{Centre for Quantum Technologies, National University of
  Singapore, Singapore}

\author{Th. Busch}

\affiliation{Department of Physics, University College Cork, Cork,
  Republic of Ireland} 

\author{V. Vedral}

\affiliation{Centre for Quantum Technologies, National University of
  Singapore, Singapore}

\affiliation{School of Physics and Astronomy, University of Leeds,
  Leeds, LS2 9JT, UK}
\begin{abstract}
  We outline an interferometric scheme for the detection of bi-mode
  and multi-mode spatial entanglement of finite-temperature,
  interacting Bose gases. Whether entanglement is present in the gas
  depends on the existence of the single-particle reduced density
  matrix between different regions of space. We apply the scheme to
  the problem of a harmonically trapped repulsive boson pair and show that while
  entanglement is rapidly decreasing with temperature, a significant
  amount remains for all interaction strengths at zero
  temperature. Thus, by tuning the interaction parameter, the
  distribution of entanglement between many spatial modes can be
  modified.
\end{abstract}

\pacs{03.67.Mn,03.75.Gg,67.10.Ba }

\maketitle

\section{Introduction}

Understanding and controlling entanglement in many-body systems is one
of the most important challenges in quantum mechanics today. The
rewards are significant and are expected to not only
lead to new insights into the properties of solid state systems and
phase transitions \cite{Amico:08}, but also to new designs for highly efficient quantum information devices. Ultra-cold bosonic gases offer an ideal arena to explore many-body entanglement, as experimentalists have at their disposal
{\sl designer} condensed matter systems whose parameters can be
controlled with unprecedented precision \cite{Bloch:08}.

Entanglement often exists naturally in the ground state of a many-body
system \cite{Audenaert:02}, where it resides between the
degrees of freedom of the particles and is a property of
the first quantised many-body wavefunction. However, in ultra-cold
gases the particles are inherently indistinguishable, which requires
the symmetization of their many-body wavefunction and means that the
Hilbert space no longer has the tensor product structure required to define entanglement. The first quantised many-body
wavefunction of indistinguishable particles may therefore contain
quantum correlations \cite{ Paskauskas:01}, but such correlations are
usually considered unable to violate a Bell inequality or process
quantum information \cite{Wiseman:03, Paskauskas:01, Dowling:06}.

Ultra-cold gases are also well described within the framework of
second quantisation, where instead of working directly with the
many-body wavefunction, one defines a complete set of field modes that
are occupied by particles.  Second quantisation therefore offers the
possibility of entanglement between modes. Entanglement is dependent
on the choice of modes, but provided the correct choice is made,
investigating entanglement between {\sl distinguishable} modes
\cite{Simon:02,Libbythesis:08,Esteve:08} circumvents the difficulties of
defining entanglement between indistinguishable particles
\cite{You:06, Murphy:07}.

To illustrate the differences between particle and mode entanglement,
let us consider two non-interacting bosons in a trap at zero
temperature.  In first quantisation, the wavefunction is the
symmetrized product, $\Psi_{12}(x,y) = \frac{1}{\sqrt{2}}
(\phi_1(x)\phi_2(y)+\phi_1(y)\phi_2(x))$, where $\phi(x)$ is the
ground state of the confining potential. No entanglement exists
between the particles, since indistinguishability forbids us from
assigning to any particle a specific set of degrees of freedom.
Conversely, in second quantisation one can define a pair of spatial
modes, $A$ and $B$, where each mode occupies half the confining
geometry.  Since both the particles are coherently distributed over
these modes, the system is described by the entangled
state, $|\psi_{AB}\rangle=
\frac{1}{2}(|20\rangle+\sqrt{2}|11\rangle+|02\rangle)$, where
$|mn\rangle=|m\rangle_A\otimes|n\rangle_B$ denotes $m$ particles in
mode $A$ and $n$ particles in mode $B$ (with $m+n=2$).

In this \paper \, we outline a scheme for the detection of bi-mode and
multi-mode spatial entanglement for a finite temperature, interacting
Bose gas of any (including unknown) particle number. We show that entanglement is detected via  the single-particle
reduced density matrix (SPRDM). We apply our scheme to the
  example of a harmonically trapped, interacting boson pair
\cite{Busch:98}, where the SPRDM also acts as a quantifier of entanglement.  We find that for all interaction strengths, entanglement between pairs of modes rapidly decreases with temperature.  While at zero temperature a significant amount of entanglement remains even in the limit of infinite interaction.   Moreover, we note that our detection scheme is also relevant to
recent proposals to observe non-locality of single particle between
spatial modes \cite{Dunningham:07,Heaney:08}.

{\begin{figure}[tb]
\begin{center}
\includegraphics[width=\linewidth] {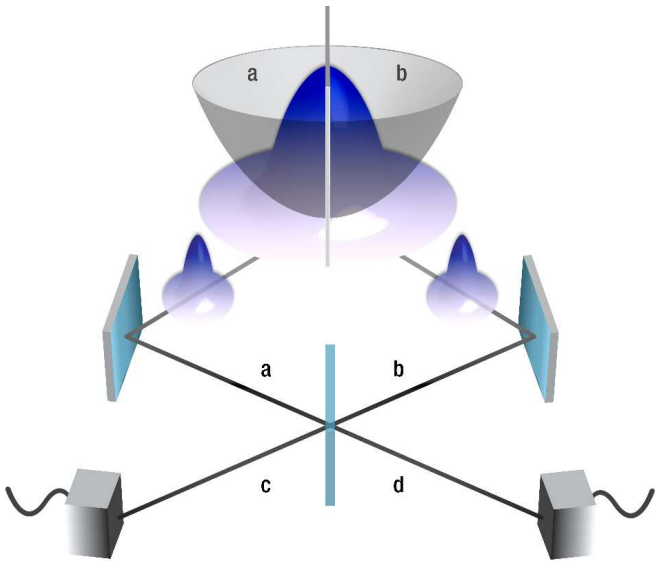}
\end{center}
\caption{ Schematic showing a trapped wavefunction split into two
  spatial modes, $a$ and $b$. The modes are combined at
  a $50/50$ beamsplitter and the particles in the output modes $c$ and
  $d$ are counted.  }
\label{fig:Schematic}
\end{figure}}

\section{Detection scheme} 

The correlations of entanglement are locally
basis independent, so that one needs to measure each mode in at least
two bases in order to differentiate them from classical correlations.
While a superselection rule that forbids coherent superpositions of
eigenstates of different mass \cite{Bartlett:07} seems to rule such measurements out for any atomic system, recent work
has shown that such measurements are theoretically possible
\cite{Heaney:08}.  Moreover, it has recently been predicted that this natural mode entanglement of massive particles can be used as a resource for quantum communication \cite{Heaney:09}.
However, in practice it will be difficult to locally
manipulate, i.e.~rotate, the spatial modes. In the following we will show that spatial entanglement can also be
detected and quantified by making \emph{global} operations on the
modes.

 Let us consider a gas in a confining geometry (see Fig.~\ref{fig:Schematic}) which is mathematically, but not
necessarily physically, divided into two, non-overlapping spatial
modes, $a$ and $b$.  The field operators,
$\hat\psi_i^{\dag}=\int_id\vec{x}\,g^*(\vec{x})\hat\psi^{\dag}(\vec{x})$
and $\hat\psi_i=\int_id\vec{x}\,g(\vec{x})\hat\psi(\vec{x})$, create
and destroy particles in mode $i=a,\,b$, where
$\int_i|g(\vec{x})|^2d\vec{x}=1$ ensures that the commutation relations,
$[\hat\psi_i,\hat\psi^{\dag}_j]=\delta_{ij}$, are satisfied. The
quantity, $g(\vec{x})$, specifies how the set of points in a spatial
mode are averaged over.  To generate interference, the particles in
the two spatial modes, $a$ and $b$, are mixed at a 50:50 beamsplitter,
which transforms the input modes as
$\hat\psi^{\dag}_a=\frac{1}{\sqrt{2}}(\hat\psi^{\dag}_c+\hat\psi^{\dag}_d)$
and
$\hat\psi^{\dag}_b=\frac{1}{\sqrt{2}}(\hat\psi^{\dag}_c-\hat\psi^{\dag}_d)$.
After the beamsplitting operation, the number of particles in the
output modes, $c$ and $d$, are counted and compared to the fully
separable case.  If the number of coincidences is different to the
separable case, we can conclude that there must have been entanglement
between the spatial modes.

 A bosonic gas of fixed particle number, $N$ which is in a
  fully separable state w.r.t.~the spatial modes $a$ and $b$ can be
  written as
\begin{equation}
\label{Eq:SepState}
    \hat\rho_{sep} = \sum_{n=0}^N p_{n} 
                     |n\rangle\langle n|_a\otimes|N-n\rangle\langle N-n|_b,
\end{equation}
where $\sum_{n}p_{n}=1$. The beamsplitter transforms this state such
that if the total number of particles is even, one
  detects the same number of particles in each of the output modes,
  i.e.~$\Delta N=|N_c-N_d|=0$. For an odd number of particles an
ensemble average leads to the same result \cite{ParticleUncertainty}.

Conversely, if the initial state of fixed particle number
  is of an arbitrary form, $\hat\rho$, the difference in particle
  numbers detected in the modes $c$ and $d$ may be non-zero due to
  entanglement between the modes. 
\begin{equation}
  \Delta N=\left|\tr\left[\hat\psi_c^{\dag}\hat\psi_c\hat\rho\right]
                -\tr\left[\hat\psi_d^{\dag}\hat\psi_d\hat\rho\right]\right|
          =2|\epsilon_{ab}|,
\end{equation}
where
\begin{equation}
\label{eq:measure}
  \epsilon_{ab}=\tr\left[\hat\psi_a^{\dag}\hat{\psi}_b\;\hat\rho\right]
              =\int_{a}d\vec{x}\int_{b}d\vec{x}'g(\vec{x})g^*(\vec{x}')
               \rho^{(1)}(\vec{x},\vec{x}').
\end{equation}
We have derived the above result using the Fourier decomposion of the
field operators, $\hat\psi^{\dag}(\vec{x})$ and $\hat{\psi}(\vec{x})$
in terms of the momentum modes, $\phi_k(\vec{x})$, as
$\hat\psi^{\dag}(\vec{x})=\sum_k\phi_k^*(\vec{x})\,\hat a_k^{\dag}$
and likewise for $\hat\psi(\vec{x})$.  When $\epsilon_{ab}$ is
non-zero, the state $\hat\rho$ is different to the separable case and
is therefore necessarily entangled w.r.t. the bi-modal split into $a$ and $b$.

The quantity $\epsilon_{ab}$ is given by the off-diagonal
  elements of the SPRDM which is defined as
$\rho^{(1)}(\vec{x},\vec{x}')=\sum_{\vec{k}}n_{\vec{k}}
\phi_{\vec{k}}(\vec{x})\phi^*_{\vec{k}}(\vec{x}')$. Here $n_{\vec{k}}$
is the number of bosons that occupy the $k$-th momentum mode.  The
SPRDM is a one-body correlation function, which characterises
important coherence properties of a many-body system \cite{Penrose:56}
and its off-diagonal elements are related to the
visibility of interference fringes in a two slit experiment
\cite{Bloch:00}.

Our scheme therefore applies to systems whose
correlations from a basic group of one particle \cite{Yang:62}. States
of the form, $|20\rangle+|02\rangle$, whose correlations are second
order are not detected by the SPRDM.  Such states require careful
engineering of bosonic systems \cite{Folling:07}, although they may
form naturally in fermionic systems, i.e.~ Cooper pairs in
superconductors. Our scheme is applicable to all bosonic gases trapped
in orthodox geometries.

Next, let us show that $\epsilon_{ab}$ can also be used
to {\sl quantify} mode entanglement for certain systems, one of which
is the boson pair model discussed below. For this
  $\epsilon_{ab}$ must fulfil three basic criteria \cite{Vedral:97}:
(i) $\epsilon_{ab}$ is zero when the state is separable, (ii)
$\epsilon_{ab}$ is invariant under local unitary operations and (iii)
$\epsilon_{ab}$ does not increase under local general measurements and
classical communication (LGM+CC). The validity of (i) is shown above
and (ii) is guaranteed since the trace is basis independent. One can
prove (iii) as follows.  To implement LGM+CC, the two spatial modes,
$a$ and $b$, are each coupled to a local environment by a general
completely positive map.  The environments are allowed to communicate
classically ad infinitum and the total number of particles in the gas
and environment is fixed. One can then show that $\epsilon_{ab}$ does not increase under
LGM+CC.

The above scheme can also analyse multi-mode entanglement, i.e.~the
simultaneous entanglement of more than two modes.  In the following we
use the different notions of separability discussed in
\cite{Shchukin:06}.  A general $M$-mode state is fully separable and
contains no entanglement if it is a convex combination of states for
each mode, $ \hat{\rho}_{sep(M)} = \sum_i p_i
\hat{\rho}_i^{(1)}\otimes\cdots\otimes\hat{\rho}_i^{(M)}$. Here each
composite state, $\hat{\rho}_i^{(j)},$ corresponds to a single spatial
mode with a fixed number of particles.  On the other hand,
entanglement may be present between certain subsets of spatial modes,
for which one can define a $p$-separable state, $\hat{\rho}_{sep(p)}
=\sum_i p_i \hat{\rho}_i^{(1)}\otimes\cdots\otimes\hat{\rho}_i^{(p)}$,
where $p\leq M$ with equality when the state is fully separable as
above.  When the composite state, $\hat{\rho}_i^{(j)},$ describes more
than one spatial mode, the spatial modes contained within
$\hat{\rho}_i^{(j)}$ are necessarily entangled otherwise
$\hat{\rho}_i^{(j)}$ would be written as a product of states for the
individual modes,
$\hat{\rho}_i^{(j)}=\hat{\rho}_i^{(j_1)}\otimes\cdots\otimes
\hat{\rho}_i^{(j_4)}\otimes\cdots$, and the overall state of the
system would be $\sigma$-separable, where $p<\sigma\leq M$
\cite{Shchukin:06}.

To determine whether a given $M$-mode state contains multi-mode
entanglement, one can check a bi-partite entanglement criterion
between all $2^{M-1}-1$ unique divisions of the system into two
\emph{blocks}, $A$ and $B$. Depending on
which pairs of blocks are found to be separable one can conclude that
the state has entanglement between different subsets of spatial modes.
Here the bipartite entanglement criterion is
$\epsilon_{ AB}=\tr[\hat\Psi_{ A}^{\dag}\hat\Psi_{ B}\hat\rho]$ of
eq.~\eqref{eq:measure}, i.e.~the SPRDM between two blocks of spatial
modes, ${ A}$ and ${ B}$. The field operators, $\hat\Psi_{{
    X}}^{\dag}$ and $\hat\Psi_{{ X}}$, for blocks of spatial modes are
defined as $\hat\Psi_{X}^{\dag}=\sum_{i\in {
    X}}c_i\,\hat{\psi}^{\dag}_i$ and $\hat\Psi_{ X}=\sum_{i\in {
    X}}c_i^*\,\hat{\psi}_i$, where ${ X} = { A}, { B}$ and $\sum_{i\in
  { A}}|c_i|^2=1$ ensures that the commutation relations,
$[\hat{\Psi}_{ X},\hat\Psi_{ Y}^{\dag}]=\delta_{{ X}, { Y}}$, are
satisfied.

We now describe how entanglement changes for the (i) fully separable (ii) the $p$-separable and (iii) the fully entangled
states: (i) The fully separable state, $\hat{\rho}_{sep(M)}$, admits
no interference between all $2^{M-1}-1$ partitions into blocks,
i.e. $\epsilon_{{ AB}}=0$ for all ${ A}$ and ${ B}$.  (ii) A
$p$-separable state, $\hat{\rho}_{sep(p)}$ where $p<M$, admits no
interference for $2^{p-1}-1$ partitions into blocks from the total of
$2^{M-1}-1$ partitions, i.e.~$\epsilon_{{AB}}=0$ for $2^{p-1}-1$
choices of $A$ and $B$.  (iii) A fully entangled state,
$\hat\rho=\sum_ip_i\hat\rho_i$, admits interference for all
$2^{M-1}-1$ choices of blocks, i.e.~$\epsilon_{AB}\neq 0$ for all
${A}$ and ${B}$.  The fully entangled state, $\hat{\rho}$, necessarily
contains some form of multi-mode entanglement, since otherwise there
would exist a decomposition such that $\hat\rho$ is $p$-separable.  

{\begin{figure}[tb]
    \includegraphics[width=\linewidth]{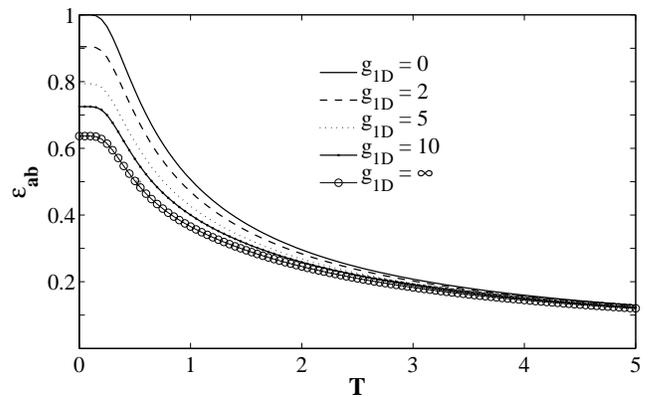}
    \caption{Amount of entanglement between two spatial regions
      occupied by a boson pair as specified in the text
      as a function of temperature. The values of the dimensionless
      interaction parameter increase from the top curve to the bottom
      curve as $g_{1D}=0,2,5,10,\infty$. Temperature is
        scaled in units of $\hbar\omega/k_{B}$.}

\label{fig:thermal}
\end{figure}}

\section{Boson pair model}

 In the following we will apply our scheme
  to the physically realistic model of a harmonically trapped pair of
  ultracold, interacting, bosonic atoms in effectively one
  dimension. The Hamiltonian of such a system is given by
\begin{equation}
  \label{eq:Hamiltonian}
  \hat H=\sum^2_{i=1}\left(-\frac{1}{2}\frac{d^2}{dx^2_i}
    +\frac{1}{2}x^2_i\right)
  +g_{1D}\delta(|x_i-x_j|),
\end{equation}
 where all lengths are scaled in units of the ground state
  size and all energies in units of the harmonic frequency. The one
dimensional coupling constant, $g_{1D}$, is related to the three
dimensional $s$-wave scattering length by $ g_{1D}=\frac{\hbar^2
  a_{3D}}{ma_\perp} \left(a_\perp-Ca_{3D}\right)^{-1}$, where $C$ is
the constant, $C=1.4603$ \cite{Olshanii:98}.  This Hamiltonian can be
decoupled by moving into the centre of mass and relative coordinate
frames labelled by $X$ and $x$, respectively, \cite{Busch:98} and the
two-body wavefunction can subsequently be written as
$\Psi_{n,\nu}(x_1,x_2)=\psi_n(X)\psi_{\nu}(x)$. The eigenvalues for
the centre of mass motion are given by $\epsilon_n^\text{com}
=(n+\frac{1}{2})$ for $n=0,1\dots$, with corresponding eigenstates,
$\psi_n^\text{com}(X)=\mathcal{N}_n H_n (X) e^{-\frac{X^2}{2}}$. Here
$\mathcal{N}_n$ is the normalisation constant and $H_n(X)$ are the
Hermite polynomials. For the relative motion, the single particle
eigenstates are $ \psi_\nu^\text{rel} (x) = \mathcal{N}_\nu\;
e^{-\frac{x^2}{2}} U \left( \frac{1}{4} - \frac{E_{\nu}}{2},
  \frac{1}{2}, x^2 \right)$, where $\nu = 0, 2, 4 \ldots$ and the
$U(a,b,z)$ are the confluent hypergeometric functions. The
corresponding eigenenergies, $E_\nu$, are determined by the roots of
the implicit relation $-g_{1D} = 2 \frac{\Gamma\left(
    -\frac{E_{\nu}}{2} + \frac{3}{4}\right)}{\Gamma\left(
    -\frac{E_{\nu}}{2} + \frac{1}{4}\right)}\;$ \cite{Busch:98}. The
kernel of the density operator in position representation is
$\rho_{n\nu}(x,x',x_{2})=\Psi^{\ast}_{n,\nu}(x,x_2)\Psi_{n,\nu}(x',x_2)$,
with the SPRDM defined as $
\rho^{(1)}_{n\nu}(x,x')=\int_{-\infty}^{+\infty}\rho_{n\nu}(x,x',x_{2})dx_2\;.$
The SPRDM in thermal equilibrium is given by
$\rho^{(1)}(x,x')=\sum_{n}^{\infty}\sum_{\nu}^{\infty}
P_{n\nu}\rho^{(1)}_{n\nu}(x,x'),$ where $P_{n\nu}$ is the Boltzmann
weight, $P_{n\nu}=\frac{1}{Z}\exp({\frac{-E_{n\nu}}{k_{b}T}})$,
$k_{b}$ is the Boltzmann constant and $Z$ is the partition function.

\section{  Bi- and multi-mode entanglement results } 

We define two equal length modes,
$a$ and $b$, and calculate their spatial entanglement for the boson
pair model as a function of temperature and interaction strength using
eq.~\eqref{eq:measure}. The results are displayed in
Fig.~\ref{fig:thermal}. With increasing temperature the gas becomes a
statistical mixture of momentum modes, $\phi_k(x)$, which destroys the
fixed relative phase between the spatial modes and consequently the
entanglement between them is severely depleted. The presence of
interaction also degrades the quality of the entanglement at $T=0$,
but a significant amount persists even in the Tonks-Girardeau limit,
of impenetrable bosons, $g_{1D}=\infty$. At first this may seem
surprising, since in the Tonks-Girardeau limit the system
can be mapped onto an ideal fermionic atom pair \cite{Girardeau:60},
for which one may not expect any entanglement to be
present. However, only the local properties of the bosons
  become identical to free spin-polarised fermions due to the
  Bose-Fermi mapping and therefore entanglement is not affected.
{\begin{figure}[tb]
    \includegraphics[width=\linewidth] {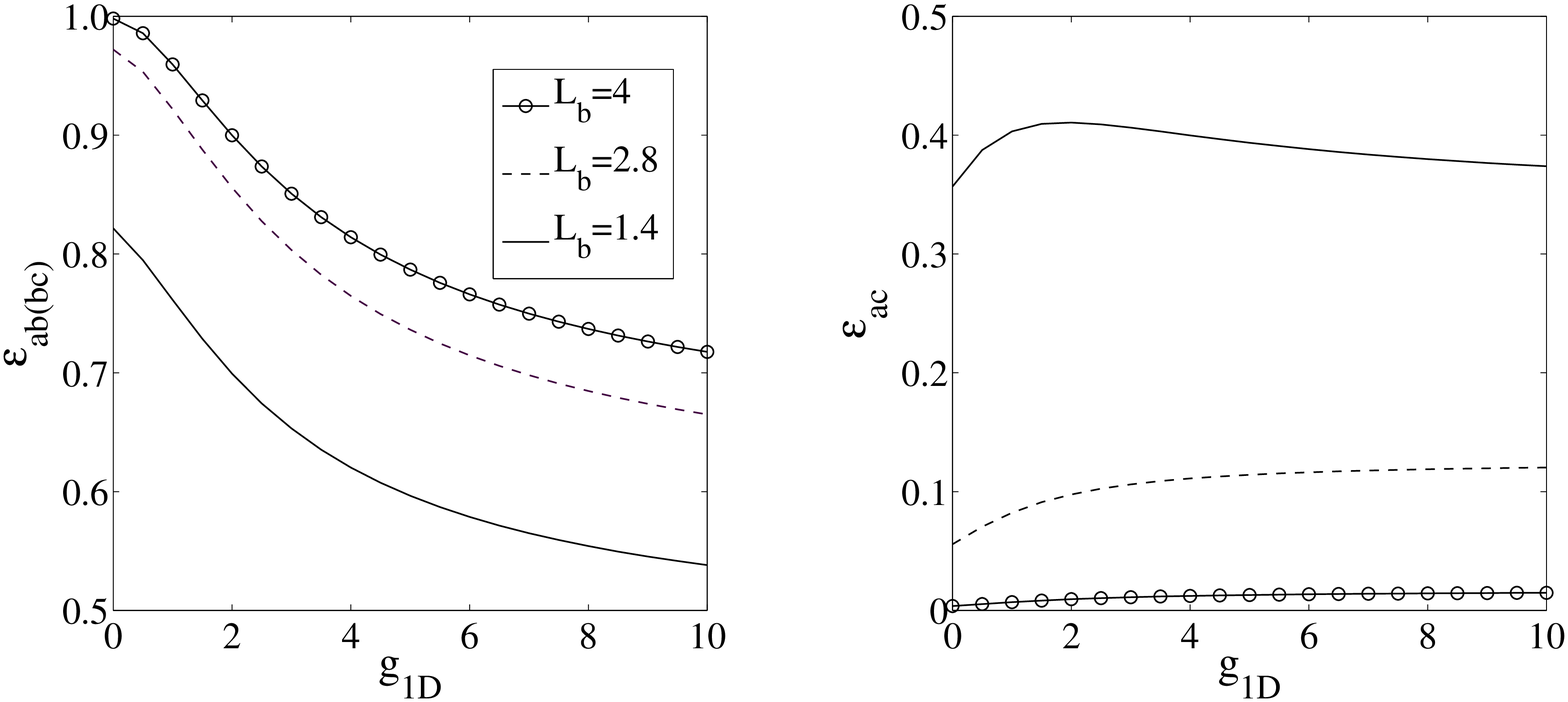}
    \caption{ The system is split into three spatial modes, a,b and
      c. Mode a is defined as the region $-\infty<x<-\frac{L_{b}}{2}$,
      mode b as ${-\frac{L{b}}{2}<x<\frac{L_{b}}{2}}$ and mode c as
      $\frac{L_{b}}{2}<x<\infty$.  Entanglement is investigated
      between neighbouring modes, $a$ and $b$ ($b$ and $c$) (left hand
      side) and also between the outer two modes, $a$ and $c$ (right
      hand side) at $T=0$ as interaction is varied. Three different
      central mode lengths, $L_{b}=4, 2.8, 1.4$ are taken.}
\label{fig:3modes}
\end{figure}}

We also examine pairwise entanglement in this model.  For this space is
divided into three modes, $a$, $b$ and $c$, and we check entanglement
between all pairs, $ab$, $bc$ and $ac$, as a function of the
interaction strength at zero temperature.  The results are shown in
Fig.~\ref{fig:3modes}, where three different central mode lengths,
$L_{b}$, are considered. The left graph of Fig.~\ref{fig:3modes} shows
entanglement between neighbouring modes, $ab$ (and for symmetry
reasons $bc$). Here entanglement decreases as the interaction strength, $g_{1D}$, increases for all lengths
studied. The right hand side graph of Fig.~\ref{fig:3modes} shows the
entanglement between outer modes, $a$ and $c$. The entanglement
increases initially for all mode sizes as the interaction strength is
varied. For the central mode size of $L_{b}=1.4$
entanglement however is not a monotonic function of $g_{1D}$ and
decreases after the initial rise. For $L_{b}=2.8$ and $4$ this is not the case. The behaviour of entanglement in both
graphs can be explained by the delocalisation of the bosons away from
the centre of the trap due to greater interaction strengths. Stronger
interaction therefore increases the correlation length (i.e. the
entanglement between the outer two modes), yet decreases the overall
coherence of the sample.

We also check the existence of multi-mode entanglement in our model. For this we have divided our two-particle wavefunction
  into $M$ modes and calculated $\epsilon_{AB}$ for the $2^{M-1}-1$
  unique partitions of these modes into two blocks, $A$ and $B$.
Recall that, $\epsilon_{AB}$, must be non-zero for all possible block
combinations for the state to be fully entangled.  We have found that
at zero temperature the system possesses full multi-mode entanglement,
i.e.~$\epsilon_{AB}\neq 0$ for all $A$ and $B$, for a wide range of
values of the interaction parameter, provided that the set of modes
are defined within the coherence length of the sample.

Here we have used the repulsively, interacting boson pair model to illustrate our scheme, since it is analytically tractable and contains all the essential features of larger models.  Our scheme to detect bi- and multi-mode entanglement can also be applied to Bose gases with a greater number of particles (for all interaction strengths and temperatures), but qualitatively the results will remain unchanged while the task of computing the single-particle reduced density matrix between modes will quickly become very demanding.  We note that bi-mode entanglement of a non-interacting Bose gas has been studied before for $N$ particles  at zero temperature \cite{Simon:02} and at finite temperatures \cite{Heaney:07}.

Our scheme can be implemented using 
currently available technologies.  Atomic beamsplitters can be
realised with optical potentials \cite{Sengstock:04} and separate modes 
can be defined using spatially selective
outcoupling techniques \cite{Bloch:00}.  To perform beamsplitting, on systems with strong interactions  one should raise a potential barrier between the desired modes  on a non-adiabatic time scale so that the coherences between the regions are not lost, (i.e. so that the gas does not enter a Mott-like state). The potential barrier should then be lowered rapidly enough so that the coupling between the wells is greater than the on-site interaction energy of the wells.  
For instance, one could envisage a setup where the trap is swiftly modulated from harmonic to double-well potential and is then switched off and the sample left to interfere. The entanglement can be inferred according to eq. \eqref{eq:measure} from the visibility of the resulting interference fringes, see for example \cite{Bloch:00, Hofferberth:07}, where mode entanglement between regions of space has already been measured indirectly.

\section{Conclusions}

We have outlined a scheme that detects bi-modal and
multi-modal spatial entanglement of cold bosonic gases using simple atom-optic techniques.  
We show that spatial entanglement is detected by the SPRDM between different modes, a quantity that is related to  the visibility of routinely measured interference fringes. We have
demonstrated our scheme using the model of a harmonically trapped
boson pair and have found the existence of bi- and multi-mode
entanglement within the gas. For all interaction strengths, increasing temperature rapidly degrades the amount of entanglement, but at zero temperature entanglement still remains, even in the
presence of strong interactions. Therefore, we have shown that the ability to vary the particle
interaction strength allows one to engineer the distribution of
entanglement over the trap, thus allowing a tunable source of entanglement.

{\it Acknowledgements} - The authors would like to thank J. Anders, P. Turner, W. Son. and M. Paternostro for valuable discussions. JG and TB acknowledge funding from
Science Foundation Ireland, project number 05/IN/I852].  LH and VV are
funded by the National Research Foundation (Singapore) and the
Ministry of Education (Singapore).


\begin{thebibliography}{99}
\bibitem{ParticleUncertainty} It is worth noting that this criterion
  means that the exact number of particles (which tends to be unknown
  in cold atom experiments and varies between different experimental
  runs) does not matter.
\end{thebibliography}
\end{document}